\documentclass{article}\usepackage{natbib,emulateapj}
\newcommand{\beq}{\begin{equation}}
\newcommand{\eeq}{\end{equation}}
\newcommand{\bea}{\begin{eqnarray}}
\newcommand{\eea}{\end{eqnarray}}
\newcommand{\rem}[1]{ }
\newcommand{\pd}[1]{\partial_{#1}}
\newcommand{\ppd}[3]{\frac{\partial^{#1} {#2}}{\partial {#3}^{#1}}}
\begin{document}
\title{The Structure of Self-Gravitating Polytropic Systems with $n$ around 5.}

\author{Mikhail V. Medvedev\altaffilmark{1} and George Rybicki}

\affil{Harvard-Smithsonian Center for Astrophysics, 60 Garden Street,
Cambridge, MA 02138}

\altaffiltext{1}{Presently at the Canadian Institute for Theoretical
Astrophysics, University of Toronto, 60 St. George Street, Toronto,
ON, M5S 3H8, Canada; medvedev@cita.utoronto.ca; 
http://www.cita.utoronto.ca/$\sim$medvedev;
also at the Institute for Nuclear Fusion, RRC ``Kurchatov
Institute'', Moscow 123182, Russia}

\begin{abstract}
We investigate the structure of self-gravitating polytropic stellar systems. 
We present a method which allows to obtain approximate analytical solutions, 
$\psi_{n+\epsilon}({\bf x})$, of the nonlinear Poisson equation with the 
polytropic index $n+\epsilon$, given the solution $\psi_n({\bf x})$ with the 
polytropic index $n$, for any positive or negative $\epsilon$ such that 
$|\epsilon|\ll1$. Application of this method to the spherically symmetric 
stellar polytropes with $n\simeq5$ yields the solutions which describe 
spatially bound systems if $n<5$ and the formation of a second core if $n>5$.
A heuristic approximate expression for the radial profile is also presented.
Due to the duality between stellar and gas polytropes, our results are valid
for gaseous, self-gravitating polytropic systems (e.g., molecular clouds) 
with the index $\gamma\simeq 6/5$. Stability of such systems and observational 
consequences for both stellar and gaseous systems are discussed.
\end{abstract}

\section{Introduction}

Galaxies, star clusters, and other stellar systems are stationary 
assemblages of stars bound by a self-consistent, mean gravitational potential. 
Their structure is determined by the nonlinear Poisson equation, which, in most
cases, may be solved only numerically. Therefore, any analytical study acquires
special interest. 

The simplest model of a gravitationally bound stellar system is a spherically
symmetric polytrope with a power-law distribution function of stars. The
gravitational potential of a polytropic sphere is determined by the
Lane--Emden equation:
\beq
\frac{1}{r^2}\frac{d}{d r}\left(r^2\frac{d}{d r}\psi(r)\right)
=-\psi^n ,
\label{LE}
\eeq
where $r$ is the dimensionless spherical radius, $\psi(r)$ is the dimensionless
potential, and $n$ is the polytropic index (see \citealp{BT} for more
discussion). The density is found from $\rho=c_n\psi^n$, 
where $c_n$ is a constant which depends on $n$.

The stellar (collisionless) polytropes are related to
hydrostatic equilibria of self-gravitating spheres of a polytropic gas with
the equation of state $P=K\rho^\gamma$, where $\gamma=1+\frac{1}{n}$, $P$
is the gas pressure, and $K$ is a constant. The gas polytropes with 
$\gamma<6/5$ (i.e., $n>5$) are usually used to model the structure of molecular 
clouds which has a major effect on the star formation that occurs within them. 
Note that the $\gamma\to1$ ($n\to\infty$) limit corresponds to the case of an 
isothermal gas. The gravitationally stable polytropes\footnote{
	Locally adiabatic polytropes with polytropic indexes $\gamma>6/5$ 
	are stable if their adiabatic indexes $\gamma_{\rm ad}>4/3$;
	globally adiabatic polytropes are unconditionally stable for
	$\gamma=\gamma_{\rm ad}>4/3$, see discussion below and, e.g., 
	\citet{McKH99}.}
with $\gamma>6/5$ (i.e., $n<5$) often serve as models of stars, because 
they can have a zero pressure boundary at a finite radius --- the star surface. 
Interestingly, the Lane-Emden equation appears also as an evolution equation 
for a scalar field in some versions of the quintessence model \citep{LS99}.
Hence the present study is also of practical interest.

There are two special cases for which
nonsingular analytical solutions of equation (\ref{LE}) are known, namely 
$n=1$ and $n=5$. We performed a study of (\ref{LE}) using a powerful 
mathematical technique --- the Lie group analysis of differential equations 
\citep{Ibragimov}. This cumbersome analysis, the details of which are 
omitted from this publication, yields that, for any $n$ except $n=1$ and $5$,
equation (\ref{LE}) is non-integrable in a closed form.\footnote{
	Strictly speaking, equation (\ref{LE}) has a zero-dimensional Lie 
	algebra, meaning that
	there is no a point transformations group admitted by (\ref{LE}).
	This does not prove, however, the global non-integrability. A
	remarkable counter-example is a class of solitonic nonlinear
	differential equations which ``became'' integrable after the relation 
	of these equations to the B\"acklund transformations group have been 
	discovered. In most cases, however, the absence of a Lie algebra
       	often means nonintegrability of a differential equation.}
When $n=1$, equation (\ref{LE}) becomes the linear Helmholtz equation, and 
when $n=5$, the solution has been discovered by \citet{Schuster1883} and takes 
a simple form:
\beq
\psi(r)=\left(1+\frac{1}{3}r^2\right)^{-1/2} .
\label{psi5}
\eeq
For other values of $n$, the general solution may be found only numerically.

The present study is motivated by a very peculiar dependence of the radial 
profiles of the gravitational potential and density on the value of $n$ when
$n\simeq5$. Several solutions of equation (\ref{LE}) with the ``core''
boundary conditions (i.e., $\psi(0)=1,\ \pd{r}\psi(0)=0$, where 
$\pd{r}\equiv\ppd{}{}{r}$) are represented in Figure \ref{fig1}. A small 
change in $n$ results in a drastic modification of the profiles. Solutions 
with $n>5$ exhibit a multiple core structure, --- the fact which has probably 
been known but never acknowledged in the literature. In this solutions,
the regions where $\psi$ falls as a power-law are interchanged with flat
cores where $\psi\simeq\textrm{constant}$. For $n<5$, the
solutions are spatially bound, i.e., $\psi$ and, hence, $\rho$ vanish at a
finite radius. The radius at which the second core begins to form in the 
$n>5$ solutions is comparable to the radius at which the density vanishes in 
the $n<5$ solutions.

In this paper we develop a functional series expansion technique which allows 
the derivation of solutions with the polytropic index $n\pm|\epsilon|$, given 
the solution with the index $n$ (for any $n$ and $\epsilon,\ |\epsilon|\ll 1$),
to any accuracy in powers of $\epsilon$ and present it in \S\ref{S:FORM}. 
We apply this technique to the $n=5$ solution (\ref{psi5}) and 
{\em analytically} derive a first-order ($\propto\epsilon^1$) approximate 
solution in \S\ref{S:5}. This solution perfectly agrees
with the exact numerical $n<5$ solutions everywhere, and with the
$n>5$ solutions at radii smaller than the second core radius.
We also provide a useful approximation for $\psi(r)$ for $n\simeq5$ to
any power in $\epsilon$. These results are directly applicable to the 
polytropic gas systems with $\gamma\simeq6/5$. In \S\ref{S:STAB} we discuss 
stability of stellar polytropes. We discuss the results and make
some observational predictions in \S\ref{S:DISC}.

\section{Analytical study of polytropic solutions: formalism \label{S:FORM} } 
 
Let us rewrite equation (\ref{LE}) in a form of the nonlinear Poisson equation 
with some boundary conditions,
\bea
& &\nabla^2\psi_n({\bf x})+\psi_n^n({\bf x})=0 , \label{main} \\
& &\left.\psi_n({\bf x})\right|_S=f({\bf x}) ,  \nonumber
\eea
where $\nabla^2$ is the three-dimensional Laplace operator, $f({\bf x})$ is
some specified function, and $S$ is a two-dimensional surface. 
The subscript $n$ explicitly shows that $\psi_n({\bf x})$ is a function of
the parameter $n$, as well as three-dimensional coordinates ${\bf x}$.
Our goal is to find a solution $\psi_{n+\epsilon}({\bf x})$, given the
solution $\psi_{n}({\bf x})$ for positive or negative $\epsilon$.
For $|\epsilon|\ll1$, such a solution may be expressed as a functional
expansion in powers of $\epsilon$:
\beq
\psi_{n+\epsilon}({\bf x})=\psi_n({\bf x})
+\epsilon\left.\ppd{}{\psi_i({\bf x})}{i}\right|_{i=n}
+\frac{\epsilon^2}{2!}\left.\ppd{2}{\psi_i({\bf x})}{i}\right|_{i=n}
+\cdots .
\label{expan}
\eeq
Note, the derivative terms above are functions of coordinates ${\bf x}$ 
evaluated at a particular value of the polytropic index. Hereafter we use 
the short-hand notation for the derivatives, 
$\left.\ppd{m}{\psi_i({\bf x})}{i}\right|_{i=n}
\equiv\pd{n}^m\psi_n({\bf x}),\ m=1,2,\dots$. The derivatives $\pd{n}^m\psi_n$
are found by solving equations (\ref{der1}) and (\ref{derm}) below. 
Differentiating (\ref{main}) with respect to $n$, we obtain
\bea
& &\nabla^2(\pd{n}\psi_n)+n\psi_n^{n-1}(\pd{n}\psi_n)=-\psi_n^n\ln\psi_n ,
\label{der1}\\
& &\left.\pd{n}\psi_n({\bf x})\right|_S=0 , \nonumber 
\eea
where the last equation means that boundary conditions are independent of $n$.
Equation (\ref{der1}) is {\em linear} in the function $\pd{n}\psi_n$ and, 
hence, is easier to solve than the original (\ref{main}). Once the solution 
of a homogeneous equation 
[i.e., equation (\ref{der1}) with vanishing right hand side] 
is found, it is straightforward to find a solution of inhomogeneous 
equation (\ref{der1}), as shown in the Appendix. The quantity
$\pd{n}\psi_n$ represents the first-order correction in equation (\ref{expan}).
Higher-order terms are obtained by taking higher derivatives of (\ref{main})
and solving for appropriate functions. For instance, upon differentiating 
(\ref{main}) $m$ times, the equation for the $m$-th derivative, 
$\pd{n}^m\psi_n$, takes the form
\bea
& &\nabla^2(\pd{n}^m\psi_n)+n\psi_n^{n-1}(\pd{n}^m\psi_n) 
\nonumber\\ & &\qquad\qquad
=-H_{n,m}\left(\psi_n,\pd{n}\psi_n,\dots,\pd{n}^{m-1}\psi_n\right) , ~~~~
\label{derm}\\
& &\left.\pd{n}^m\psi_n({\bf x})\right|_S=0 . \nonumber 
\eea
Here $H_{n,m}(\cdots)$ is a function which depends on $n$, $m$, $\psi_n$, 
and all previously found functions 
$\pd{n}^k\psi_n,\ 1\le k\le m-1$. Note that the homogeneous solutions of 
equation (\ref{derm}) for all $m$ are the same and identical to the 
homogeneous solution of equation (\ref{der1}). Hence, the calculation 
of $\pd{n}^m\psi_n$ is staightforward (see Appendix).

\section{The structure of stellar polytropes with $n=5\pm|\epsilon|$ 
\label{S:5} }

Equations (\ref{expan}), (\ref{derm}) allow the solutions
$\psi_{n+\epsilon}$ to be calculated from a solution $\psi_n$.
As mentioned above, there are two cases in which $\psi_n$ is available.
For $n=1$, equation (\ref{main}) is linear and the complete three-dimensional 
solution, $\psi_1({\bf x})$, is known. For $n=5$, only the spherically 
symmetric solution, $\psi_5(r)$, has been discovered, which is given by equation
(\ref{psi5}). Here we consider the latter case and find an analytical
expression for the first-order correction.

The function $\psi_5(r)$ is given by equation (\ref{psi5}).
Equation (\ref{der1}) for $n=5$ takes the form,
\beq
\frac{1}{r^2}\frac{d}{d r}\left(r^2\frac{d}{d r}\phi(r)\right)
+\frac{5}{(1+r^2/3)^2}\,\phi(r)
=\frac{1}{2}\frac{\ln(1+r^2/3)}{(1+r^2/3)^{5/2}} ,
\label{DER1-5}
\eeq
where we again introduced a short-hand notation 
$\phi(r)=\pd{n}\psi_5=\left.\ppd{}{\psi_i(r)}{i}\right|_{i=5}$. The general
solution of this equation may be obtained if a special solution of the
homogeneous equation [i.e., equation (\ref{DER1-5}) with the zero right hand 
side] is known, as shown in the Appendix. Since $\psi_5$ is a power-law of a
rational function, we guess that a special solution of the
homogeneous equation is a function of the same kind. We have the first
homogeneous solution, 
\beq
\phi_1(r)=\frac{r^2-3}{(r^2+3)^{3/2}}
\eeq
The second homogeneous solution is obtained from equation (\ref{y2}) where 
$\Delta=1/r^2$, as follows from equation (\ref{delta}). It reads
\beq
\phi_2(r)=\frac{r^4-18r^2+9}{r(r^2+3)^{3/2}} .
\eeq
The general solution is found from equation (\ref{y}) by straightforward
integration. The constants of integration are found from boundary conditions.
First, we require the solution to be non-diverging everywhere; hence $C_2=0$.
Second, we consider the solutions with a core, i.e.,
$\psi_{5}(0)=\psi_{5+\epsilon}(0)=1$; hence, $\phi(0)=\pd{n}\psi_5(0)=0$.
The last condition yields $C_1=\sqrt{3}/12$. Finally, we have
\beq
\psi_{5+\epsilon}(r)\simeq\psi_5(r)+\epsilon\,\pd{n}\psi_5(r),
\label{soln}
\eeq
where
\bea
\psi_5(r)&=&\left(1+\frac{r^2}{3}\right)^{-1/2} ,\\
\pd{n}\psi_5(r)&=&\frac{(r^4-18r^2+9)}{16r(r^2+3)^{3/2}}\,
\textrm{arc\,tan}\!\left(\frac{r}{\sqrt{3}}\right)
\nonumber\\
& &{ } +\frac{3^{3/2}/4}{(r^2+3)^{3/2}}
\left[\frac{7r^2}{36}-\frac{1}{4}+\ln\!\left(1+\frac{r^2}{3}\right)\right]. 
~~~~~~
\eea
The approximate solution (\ref{soln}) is shown in Figure \ref{fig1} by dotted
curves for several values of $\epsilon$. The function $\pd{n}\psi_5$ is also
shown by the dashed line. It is seen that solutions with $n<5$ are very
well approximated by the first-order expansion (\ref{soln}) at all radii.
In contrast, for solutions with $n>5$, this approximation is valid for $r$ less
than or comparable to the second core radius, i.e., where the second ``plateau''
ends. To obtain the profiles at larger radii, higher-order terms in $\epsilon$ 
must be kept. The next, $\epsilon^2$-term can also be calculated 
analytically. It is, however, cumbersome and involves 
generalized hypergeometric functions and, therefore, is not presented here.
As for the higher-order corrections, the complexity of the resulting integrals 
seems to preclude further analytical treatment.

We can now estimate the inner radius of the second core as follows. The 
asymptotic value of $\pd{n}\psi_5$ at $r\gg1$ is $\pd{n}\psi_5(\infty)=\pi/32$.
The transition from a power-law decay to a flat core
occurs when the first and second terms are of comparable
magnitude, $\psi_5(r_*)\simeq\pi|\epsilon|/32$. Thus, the transition radius is
\beq
r_*\simeq\frac{32\sqrt{3}}{\pi|\epsilon|} .
\label{r*}
\eeq
Note that for $\epsilon<0$ this radius determines the size of the system,
i.e., the radius at which the density vanishes.

Without proof we suggest that the inner and outer core radii of other
cores are also inversely proportional to the powers of $\epsilon$.
Using this conjecture, we present a heuristic expression
which approximates the $n=5+|\epsilon|$ solution quite well. The expression is
best written as a continuous fraction,
\beq
\psi_{5+|\epsilon|}(r)\simeq\frac{1}{\sqrt{
1+\displaystyle{\frac{r^2/r_1^2}{
1+\epsilon^2\,\displaystyle{\frac{r^2/r_2^2}{
1+\epsilon^4\,\displaystyle{\frac{r^2/r_3^2}{
1+\cdots
}}} }}} }}\ ,
\label{fraction}
\eeq
where $r_1,\ r_2,\ r_3,\dots$ are constants independent of $\epsilon$ 
which are to be found by fitting to the
exact numerical solution with $\epsilon\to0$. For the first two radii, we use
the analytical expressions: $r_1=\sqrt{3}$ and $r_2=r_*|\epsilon|$. Truncating
the fraction at the $r_7$ term and performing the 
nonlinear fit to the exact solution for $\epsilon=.001$, we have
\bea
r_1&=&\sqrt{3}\simeq1.73, \nonumber\\
r_2&=&32\sqrt{3}/\pi\simeq17.6, \nonumber\\
r_3&\simeq&1.95\times10^2, \nonumber\\
r_4&\simeq&9.29\times10^2, \label{values} \\
r_5&\simeq&4.75\times10^3, \nonumber\\
r_6&\simeq&1.60\times10^4, \nonumber\\
r_7&\simeq&5.00\times10^5. \nonumber
\eea
The polytropes with $n=5-|\epsilon|$ are well described by the following 
approximate expression
\beq
\psi_{5-|\epsilon|}(r)\simeq
\frac{1}{\sqrt{1+r^2/3}}-|\epsilon|\,\frac{\pi}{32} ,
\label{finite}
\eeq
which is easily obtained from equation (\ref{soln}). No higher order 
corrections are needed in this case.

How do the $n>5$ stellar polytropic spheres look to an observer?
Assuming that the luminosity density is proportional to the number density 
of stars, the surface brightness profile is calculated as follows (\cite{BT}),
\beq
I(R)=I_0\int_R^\infty\frac{\rho(r)\, r\, d r}{\sqrt{r^2-R^2}} ,
\eeq
where $R$ is the radial distance from the center of an object on the plane 
of the sky and $I_0$ is a normalization. Note, if $\rho(r)$ falls too slowly 
and the integral fails to converge, an upper limit must be replaced with an 
effective tidal radius to truncate the system. Using approximations 
(\ref{fraction})--(\ref{values}) and that $\rho_n(r)\propto\psi_n^n(r)$, 
we numerically determine the surface
brightness profiles, as presented in Figure \ref{fig2}. As one can see, 
photometric observations will reveal a bright core surrounded by a much larger
and considerably fainter shell or ``halo'' throughout which the surface
brightness is almost constant. Other halos which are located at 
larger radii are perhaps too faint to be observable. 

We can also estimate the profile of the velocity dispersion of stars
$\sigma^2=\langle v^2\rangle-\langle v\rangle^2$ which is analogous 
to the sound speed in a gas, $\sigma^2(r)\propto P/\rho\propto\psi_n(r)$.
Therefore, the velocity dispersion decreases with radius and exhibits the same
multi-core structure. In particular, the faint halo is cooler than the
central core and looks nearly isothermal because 
$\sigma^2\simeq\textrm{ constant}$ thoughout.

\section{Stability of the Polytropes \label{S:STAB}}

The stability of self-gravitating polytropes depends on how the 
system responses to an adiabatic perturbation, i.e., whether 
the internal heat transfer through the system is long or short
compared to the dynamical scale \citep{McKH99}. In the first case 
--- a locally adiabatic polytrope --- the polytropic index, which 
determines the global structure, may differ from the adiabatic 
index $\gamma_{\rm ad}$, which determines thermodynamical properties. 
The second case represents a globally adiabatic polytrope in which 
$\gamma_{\rm ad}=\gamma$. 

The stability analysis of nonsingular 
polytropic spheres usually requires numerical calculations, even 
for the simplest cases (e.g., $n=5$). Here is a  brief summary of 
the results (see \citealp{McKH99} and references therein). First,
polytropes with $\gamma>4/3$ (i.e., $n<3$) are unconditionally 
stable for any value of $\gamma_{\rm ad}$. Second, locally adiabatic 
systems with $6/5<\gamma\le4/3$ are stable if $\gamma_{\rm ad}>4/3$.
A system with smaller $\gamma_{\rm ad}$, e.g., a globally
adiabatic polytrope, is stable only if it is truncated at a certain 
radius and confined by an appropriate external pressure. Third, for the systems 
with $\gamma\le6/5$, the truncation radius depends on the value of 
$\gamma_{\rm ad}$. This radius approaches infinity for 
\beq
\gamma_{{\rm ad},\infty}=\frac{32\gamma(2-\gamma)}{(6-\gamma)^2}<\gamma.
\eeq
As one can see, globally adiabatic polytropes with $\gamma<4/3$ are stable 
only if they are pressure confined (in an analogy with Bonnor-Ebert isothermal
sphere). The truncation radius follows from the solution of equation
(54) of \citet{McKH99}\footnote{
	There is a typo: in the first equality the power of $c_s$ 
	is 3, not 3/2; cf., their equation (10).}
which in our dimensionless units reads as
\beq
m_{\rm cr}\equiv\int_0^{r_{\rm cr}}4\pi r^2\rho(r)\,dr
=\left(\frac{4\pi}{n+1}\right)^{\frac{3}{2}}\mu_{\rm cr}
\left[\rho(r_{\rm cr})\right]^{\frac{3}{2n}-\frac{1}{2}}.
\eeq
Here $\mu_{\rm cr}$ is a factor determined numerically; it is close 
to unity for large $n$ and monotonically approaches $\simeq4.6$ as $n\to3$.
For $n=5$ and $r_{\rm cr}\gg r_1$ this equation yields
\beq
r_{\rm cr}=\frac{18}{\mu_{\rm cr}}\sqrt{\frac{3}{2\pi}}.
\eeq
Comparing this to equation (\ref{r*}), we see that $r_{\rm cr}\simeq r_*$
corresponds to 
\beq
\epsilon\ga\frac{16}{9}\sqrt{\frac{2}{\pi}}\mu_{\rm cr}\sim1,
\eeq
i.e., $n\sim6$. Therefore, the multiple core structure in a globally adiabatic 
system may be seen if its polytropic index is $n\ga6$. The stability analysis
of these systems requires the use of the density profiles given by
equations (\ref{soln}) or (\ref{fraction}). The accuracy of our
approximate solution degrades for $\epsilon>1$. The qualitative 
behavior, however, remains correct (note that in the limit of an 
isothermal sphere $\epsilon\to\infty$, some oscillations in the 
profile persist). On the other hand, the multiple core structure may
be seen in pressure confined, locally adiabatic ($\gamma_{\rm ad}\not=\gamma$) 
polytropic systems with $n>5$ and $\gamma_{\rm ad}$ being sufficiently
close to unity, so that $r_{\rm cr}\gg r_*$.

\section{Discussion \label{S:DISC} }

In this paper we have demonstrated that the radial structure of polytropic
self-gravitating spheres is very sensitive to the value of the polytropic
index $n$ when $n\simeq5$ and have investigated it analytically. 
We developed a method which allows one to obtain approximate solutions, 
$\psi_{n+\epsilon}$, of the Lane-Emden equation, provided the solution 
$\psi_n$ is known and $\epsilon$ is a positive or negative constant,
$|\epsilon|\ll1$. We applied this method to $n=5$ solution and obtained the
first-order (in $\epsilon$) correction analytically. The obtained solution
describes finite-size systems with $n<5$ and the formation of the second core 
in systems with $n>5$. Both effects occur at the radius 
$r_*\simeq32\sqrt{3}/\pi|\epsilon|$. We also provided a continuous fraction
approximation for the polytropes with $n>5$, equation (\ref{fraction}), and
a simple approximate formula for those with $n<5$, equation (\ref{finite}).

We make the observational predictions by calculating  numerically 
the surface brightness of a stellar polytropic system. A galaxy or a star
cluster which is described by a polytrope with $n\gtrsim5$ will be observed as 
a bright core surrounded by a very extended faint halo or shell. This halo is
cooler than the central core and looks nearly isothermal. Because of the large
spatial size and low surface brightness this halo may easily be confused with a 
background. A way to disentangle the two is to probe a correlation between the 
core radii, $r_1,\ r_2$, and (if detected) $r_3$ and the surface brightness
distribution.

Due to the duality between the stellar collisionless polytropes and the gas
polytropes, our results are immediately applicable to the latter with the 
adiabatic index $\gamma=1+\frac{1}{n}\simeq\frac{6}{5}$. Such polytropic gas
spheres are used in studies of molecular clouds and star formation. 
The solutions with $\gamma<6/5$ describe extended molecular clouds bounded by
external pressure. Equation (\ref{fraction}) describes the multi-core structure 
of a cloud with $\gamma\lesssim6/5$. The density and, hence, pressure is
almost constant in core/halo regions. Therefore, if an external pressure, 
$p_{ext}$, is comparable to the pressure in one of the halos, then even a small 
variation in $p_{ext}$ will drastically change the
radius of the cloud. Thus, a large dispersion of masses and sizes of the clouds
embedded into nearly the same environments may indicate that the gas in the
cloud has $\gamma\lesssim6/5$, although other explanations are also possible. 
The gravitationally stable polytropes with 
$\gamma>6/5$ are often used as models of stars because the pressure vanishes 
at a finite radius, --- the star's surface. If $\gamma\gtrsim6/5$, the radial 
profile is well described by equation (\ref{finite}) and the radius of a star 
is given by equation (\ref{r*}).

\acknowledgements 

We thank Ramesh Narayan for interesting discussions and an anonymous referee
for very careful reading of the manuscript and many helpful suggestions.
This work was supported in part by NASA grant NAG~5-2837 
and NSF grant AST~9820686.

\begin{appendix}
\section{General solution of equation (\ref{DER1-5}) }

Let us consider the following differential equation,
\beq
y''+f(x) y'+g(x) y=h(x),
\eeq
where prime denotes the $x$-derivative and $f(x),\  g(x)$, and $h(x)$ are
some functions. The general solution of this inhomogeneous equation reads,
\beq
y=C_1 y_1+C_2 y_2
+y_2\int\frac{h}{\Delta}y_1\,dx-y_1\int\frac{h}{\Delta}y_2\, dx,
\label{y}
\eeq
where $C_1$ and $C_2$ are constants of integration, $y_1=y_1(x)$ and
$y_2=y_2(x)$ are two linearly independent solutions of the homogeneous equation
$y''+f(x) y'+g(x) y=0$. Here also 
\beq
\Delta=e^{-\int f(x)\, dx}=y_1y_2'-y_2y_1' .
\label{delta}
\eeq
If one nontrivial solution of the homogeneous equation, $y_1$, is known, then
the second may be found from the equation
\beq
y_2=y_1\int\frac{\Delta}{y_1^2}\, dx.
\label{y2}
\eeq

\end{appendix}

\figcaption[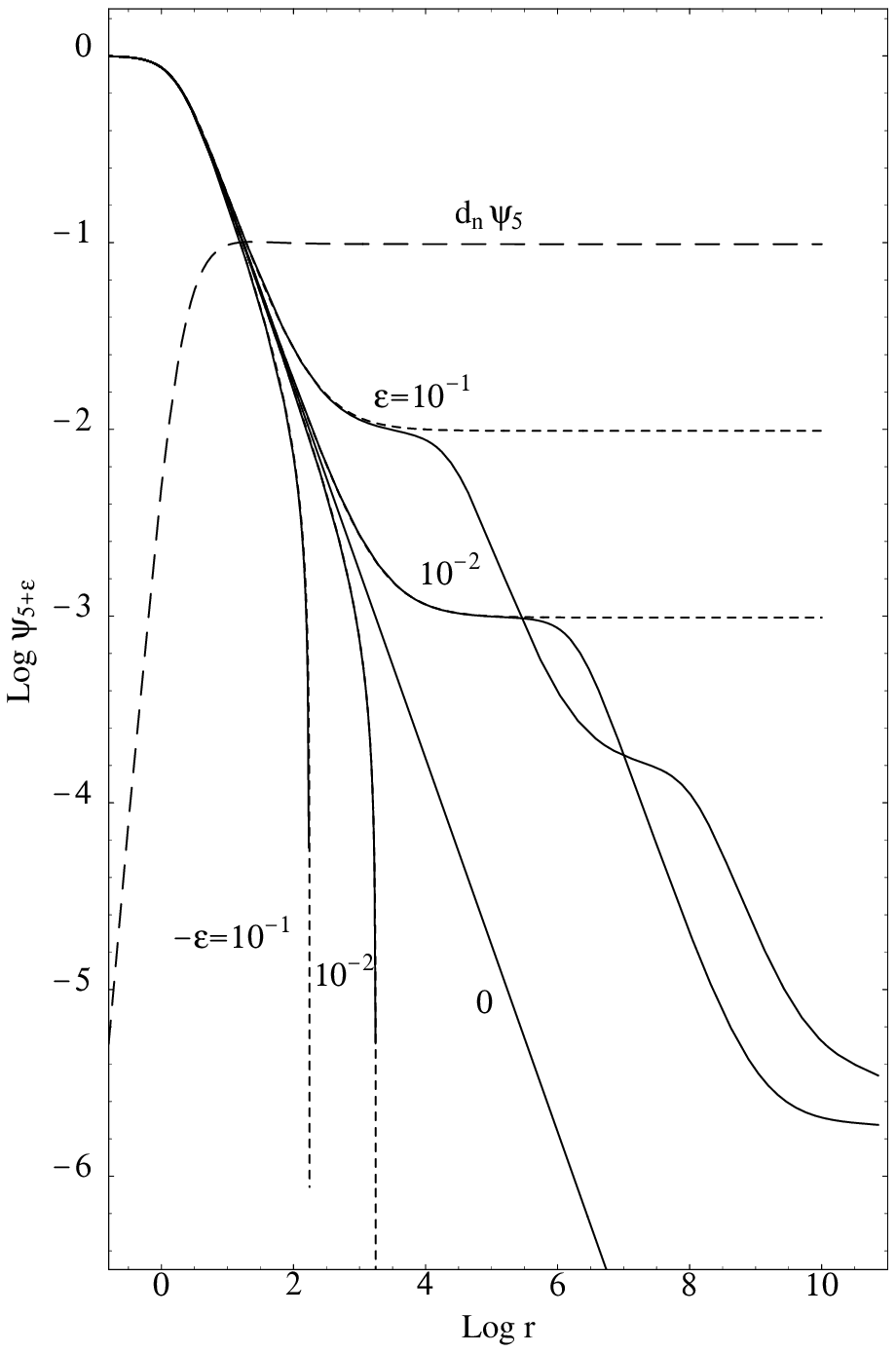]{Radial profiles of $\psi_{5+\epsilon}$ 
for several values of $\epsilon$. The solid lines are exact solutions of 
the Lane-Emden equation, the dotted lines are analytical approximate solutions,
and the dashed line is the first-order term $\pd{n}\psi_5(r)$.  \label{fig1} }

\figcaption[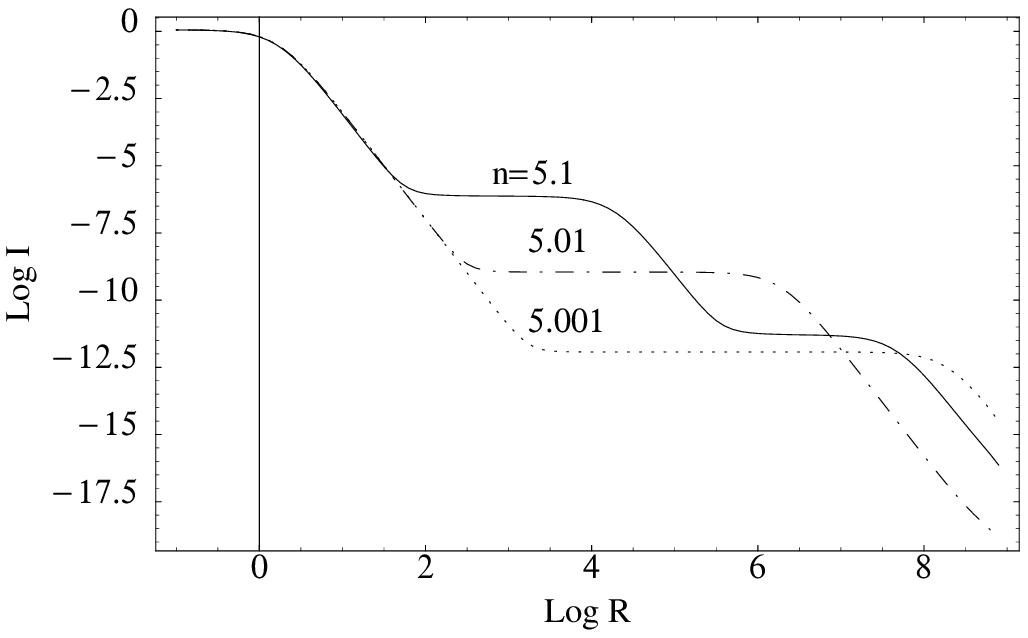]{The surface brightness profile for $n=5.1,\ 5.01$, and
$5.001$. \label{fig2} }

\vfil

\plottwo{f1.eps}{f2.eps}\\~ \\ Figs. \ref{fig1}, \ref{fig2}

\rem{
\begin{minipage}{5.5in}
\plotone{f1.eps}\\~ \\ Fig. \ref{fig1}
\end{minipage}
\newpage
\plotone{f2.eps}\\~ \\ Fig. \ref{fig2}
}

\end{document}